\documentclass{article}
\usepackage{amsmath}
\usepackage{amsfonts,amssymb,amsthm}
\usepackage{epsfig}
\newcommand{\tr}{\,\mbox{\rm tr}}
\newcommand{\PSL}{\operatorname{PSL}}
\newcommand{\RM}{\mathbb{R}}

\newtheorem{theorem}{Theorem}
\newtheorem{corollary}{Corollary}
\newtheorem{lemma}{Lemma}
\newtheorem{proposition}{Proposition}
\newtheorem{remark}{Remark}
\newtheorem{definition}{Definition}
\newtheorem{example}{Example}

\psfull

\begin{document}

\bibliographystyle{plain}
\nocite{*}

\centerline{\bf Dislocation Defects and Diophantine Approximation}

\centerline{Jared C. Bronski}
\centerline{Zoi Rapti}

\centerline{University of Illinois, \\ Department of Mathematics, }

\begin{abstract}
In this paper we consider a Schr\"odinger eigenvalue problem 
with a potential consisting of a periodic part together with a
compactly supported defect potential. Such problems arise as models in
condensed matter to describe color in crystals as well as in engineering
to describe optical photonic structures.
We are interested in studying the
existence of point eigenvalues in gaps in the essential spectrum, and in
particular in counting the number of such eigenvalues.
We use a homotopy argument in the width of the potential to count the
eigenvalues as they are created.
As a consequence of this we prove the following significant
generalization of Zheludev's theorem: the number of point eigenvalues in a
gap in the essential spectrum is exactly $1$ for sufficiently
large gap number unless a
certain Diophantine approximation problem has solutions, in
which case there exists a subsequence of gaps containing $0,1,$ or $2$
eigenvalues.
We state some conditions under which the solvability of the Diophantine
approximation problem can be established.

\end{abstract}

\section{Introduction}

Periodic media with defects have long been studied in physics and applied
mathematics. Such problems have been studied in
quantum mechanics and condensed matter physics\cite{rarity} to model
the effect of impurities on the color of  a crystal.
Since the development of photonic crystals\cite{FK,KFI,KFII}
and the ability of engineers to construct such defects
in an optical context this area has received new attention. Most of these
studies are concerned with the characterization of the spectrum
of the medium due to the presence of the defect. More precisely, since
a defect does not change the essential spectrum of the
associated operators, the main problem is to study the creation of isolated
point eigenvalues of finite multiplicity in the spectral gaps of the
unperturbed problem. 

There is a fairly extensive literature on the study of the existence and
number of bound states with energies that lie in a spectral gap.
Much of this literature \cite{DH,KFI,KFII,GS} employs some variant of
the Birman-Schwinger principle in order to count the number of
gap modes. The Birman-Schwinger principle is, in essence, a homotopy
argument in the coupling strength of the defect potential, where one tries
to estimate the number of eigenvalues below a given value of the
coupling strength.
%In \cite{GS} the authors present an alternative proof to Zheludev's theorem
%that states the finiteness of eigenvalues in a spectral gap and gives an
%exact count of them in the case of a large gap number.
Weyl asymptotics and Dirichlet-Neumann bracketing have been
used to produce asymptotic estimates for the number of defect eigenvalues
in the spectral gap of Schr\"odinger operators in the large coupling limit.
Other studies \cite{FYC} have used formal perturbation methods
to investigate the origin and distribution of defect modes.

Perhaps the closest results to the ones presented here are due to 
Zheludev\cite{ZI,ZII}, Rofe-Betekov\cite{RBI,RBII} and 
Firsova\cite{FI,FII} (see also the paper of Gesztesy and Simon\cite{GS}, which
gives a self contained derivation of these results). 
In these papers the following results are shown: 
given Schrodinger operator in one dimension 
with potential of the form $q_{per}(x) + q_{def}(x),$ where 
$q_{per}(x+a)=q_{per}(x)$ is periodic, $q_{per}\in L^1_{loc}(\mathbb{R})$, $(1+|x|)q_{def}(x) \in L^1(\mathbb{R})$ then
\begin{itemize}
\item[i] There are a finite number of eigenvalues in each gap in the essential spectrum.
\item[ii] There are at most two eigenvalues in each sufficiently large numbered gap.
\item[iii] If $\int q_{def}(x)dx \neq 0$ there is precisely one eigenvalue in each 
sufficiently large numbered gap. 
\end{itemize} 

In this paper we will be concerned with the periodic Schr\"odinger eigenvalue
problem with a compactly supported defect
\begin{align}
\label{eqn:Schro}
&-u_{xx}+q(x)\,u=E\,u ~~~~~~ x\in \RM \\
\label{eqn:Pot}
&q(x) = \left\{ \begin{array}{c}q_{per}(x) ~~~~~~x \leq 0 \\
~q_{def}(x) ~~~~~~~x\in(0,1) \\
~q_{per}(x-1) ~~~~~x\geq 1\end{array}\right.
\end{align} 
where the potential $q_{per}$ has period $a:$ $q_{per}(x+a)=q_{per}(x).$
Note that we do not necessarily assume any particular relationship
between the width of the defect (scaled for convenience to be $1$)
and the period of the potential $q_{per}(x)$ (denoted by $a$) so there
is a relative phase shift or dislocation in the potential between
$-\infty$ and $+\infty$ if $a^{-1}$ is not an integer.
In engineering applications defects are
created by cutting and layering materials with different properties, so 
the above form of the defect is actually
extremely natural from the point of view of applications\cite{Y}. This 
asymptotic phase shift in the periodic potential will 
turn out to have very interesting physical and mathematical properties, and 
does not appear to have been previously considered.

An outline of the paper is as follows: we first give a simple topological 
argument relating the number of point eigenvalues in a gap to the 
number of connected components of the intersection 
of the gap with a certain set, essentially a resolvent set associated 
with the defect. Next we give necessary Diophantine condition for 
large-numbered gaps to intersect this resolvent set. In the end we
establish the following result: Defining $\Delta q$
to be the mean energy difference between the defect and periodic potentials
\[
\Delta q = \int_0^1 q_{def}(x) dx - \frac{1}{a} \int_0^a q_{per}(x) dx
\]
there is a set $\mathcal{F}_a$ related to the distribution of errors in
Diophantine approximation of $a$ such that the following holds: if
$a \Delta q\notin\mathcal{F}_a$ then every sufficiently large numbered gap
contains exactly one eigenvalue. If  $a \Delta q \in\mathcal{F}_a$ there exists
 a subsequence of gaps related to rational approximants of $a$
which may contain $0,1$ or $2$ eigenvalues. The set $\mathcal{F}_a$
satisfies
$ \mathcal{F}_a = \mathcal{F}_{a'}$ if $a$ and $a'$ are related by an element of
the modular group $\rm{PSL}(2,\mathbb{Z})$.
The results of Zheludev, Firsova and Rofe-Betekov are analogous  to the case 
with no asymptotic phase shift, $a^{-1}\equiv 0 ~~\mod~1,$ whose equivalence 
class is the rationals. For rational $a$ the set  
$\mathcal{F}_a$ is simply the point $0$, analogous to point $(iii)$ 
above. For irrational $a$, however, the set $F_a$ is much more interesting. 
We characterize
$\mathcal{F}_a$ for certain cases, and give an explicit example of a periodic
potential and
defect having a subsequence of gaps containing exactly $2$ eigenvalues.

\section{Preliminaries}

In this section we define the important quantities for the defect eigenvalue
calculation and present some technical results on the behavior of
these quantities. Many of these are standard results which can be found in 
(for example) the text of Magnus and Winkler on Hill's equation.

\subsection{Basic Definitions and Equations}
We will usually write equation \eqref{eqn:Schro} as a system
\begin{align}
\label{eqn:ham1}
\frac{d}{dx}\left(\begin{array}{c}u \\ p\end{array}\right)
 &= J \left(\begin{array}{cc} E-q(x) & 0 \\ 0 & 1\end{array} \right)
\left(\begin{array}{c}u \\ p\end{array}\right) = J H\left(\begin{array}{c}u \\ p\end{array}\right)  \\
J &= \left(\begin{array}{cc} 0 & 1 \\ -1 & 0 \end{array}\right),\nonumber
\end{align}
where the potential $q$ is assumed to be piecewise $C^1$. This assumption 
could probably be weakened to allow Hamiltonians defined as 
quadratic forms, but we have not done this.  

It is clear from a standard Weyl sequence argument that the essential
spectrum of \eqref{eqn:Schro} is the same as the essential spectrum of the
purely periodic problem
\[
-u_{xx} + q_{per}(x) u = E u
\]
and, because the defect is compactly supported, there are no imbedded eigenvalues in the essential spectrum.
See the general discussion in Reed and Simon\cite{RS} and more specifically
Theorem 6.28
in Rofe-Beketov and Kholkin\cite{RBK}, which applies immediately to
this case. Thus we need only understand the
presence of point spectrum in the resolvent set of the periodic problem -
the spectral gaps. Let us recall some standard definitions and facts 
from Floquet theory.

We define $M(E)$, the monodromy matrix for the periodic problem, to be
\begin{equation}
M(E) = U(E,a),
\end{equation}
where the fundamental solution matrix $U(E,x)$ satisfies
\begin{align}
&U_x = J \left(\begin{array}{cc}E-q_{per}(x) & 0 \\ 0 & 1 \end{array}\right)U = J H_{per}(E,x) U \\
&U(E,0) = I. \nonumber
\end{align}
where $I$ is the identity matrix.
The monodromy $M(E)$ satisfies the identity $M^t J M = J$ (i.e. $M(E)$ is a symplectic matrix). The Floquet discriminant associated to the periodic 
problem is defined to be 
\[
k(E) = \tr(M(E))
\]
The band edges $\{E_j\}_{j=0}^\infty$ are the 
roots of $k^2(E)=4$ in increasing order and respecting multiplicity. 
The band edges  $E_k$ are
periodic points ($k(E_j)=2$) if $j \equiv 0,3 \mod 4$ and are
antiperiodic points ($k(E_j)=-2$) if  $j \equiv 1,2 \mod 4.$
An energy $E$ is said to belong to a band or a gap if $k^2(E)<4$
or $k^2(E)>4$ respectively. The bands are given by intervals 
of the form 
\[
k^2(E) < 4 ~~~~~~~~E \in (E_{2j},E_{2j+1})
\]
and the gaps by the intervals of the form
\[
k^2(E) > 4 ~~~~~~~~E \in (E_{2j-1},E_{2j}) = \mathcal{G}_j.
\]
The band edges satisfy $E_{2j}<E_{2j+1}\leq E_{2j+2}$, so that the bands 
are never empty, although the gaps may be so. The sign of the 
derivative of the Floquet 
discriminant $k'(E)$ is fixed on bands: $k'(E)\neq 0$ for energies 
in a band, $E \in (E_{2j},E_{2j+1})$, and it can only vanish at a band edge 
if $E_{2j+1}= E_{2j+2}$ and the gap is empty. An empty gap is referred to as a 
double point, and the monodromy matrix $M(E)$ is a multiple of the 
identity if and only if $E$ is a double-point. The slope of the Floquet 
discriminant in a band can be interpreted as the Krein signature\cite{K,ys,BR} 
of the eigenvalues of $M(E)$, although we will not exploit this here. 
The essential spectrum is the 
union of the bands and band edges:
\[
\sigma_{\bf ess} = \cup_{j=0}^\infty [E_{2j},E_{2j+1}]
\] 
For energies in a gap, $E\in(E_{2j-1},E_{2j})$, the monodromy matrix $M(E)$ has two
distinct real eigenvalues,
\[
M(E) v_\pm = \lambda_\pm v_\pm ~~~~~~~~|\lambda_-|<1<|\lambda_+| \\
\]
 one of modulus greater than one and one of modulus
less than one. The resolvent set of the periodic problem
\[
\mathcal{R}_{per} = \cup_{j=0}^\infty (E_{2j-1},E_{2j})
\]
will prove important. Here we use the convention that $E_{-1}=-\infty$
so the zeroth gap is the semi-infinite one $(-\infty,E_0)$.

Similarly $N(E,x)$ will denote the fundamental solution matrix associated to the defect 
\begin{align}
&N_x = J  \left(\begin{array}{cc}E-q_{def}(x) & 0 \\ 0 & 1 \end{array}\right) N = J H_{def}(E,x) N \label{eqn:fundsol} \\
&N(E,0) = I. \nonumber
\end{align}
We will define the Floquet discriminant for the defect to be 
\[
k_{def}(E,x) = \tr\left(N(E,x)\right)
\]
We will refer to a point $(E,x)$ as belonging to a band, band-edge or gap 
for the defect problem
if $k_{def}^2(E,x)<4$, $k_{def}^2(E,x)=4$,
or $k_{def}^2(E,x)>4$ respectively. 
The eigenvectors of $N(x,E)$ in a gap 
will be denoted $w_\pm:$ 
\[
N w_\pm = \tau_\pm w_\pm ~~~~~~|\tau_-|<1<|\tau_+|.
\]

We will often work with $N(E,1)$, the fundamental matrix evaluated at the
end of the support of the defect potential, which with some small
abuse of notation we refer to as the monodromy map for the defect.
This will be denoted by $N(E) = N(E,1)$.
We will also
have occasion to fix either $x$ or $E$, and consider the
the band-gap structure as the other quantity varies. In such
cases we will refer to the gaps as being ``in $x$'' or ``in $E$''
as appropriate.

For the remainder of this discussion we consider a fixed energy $E\in\mathcal{G}_j$ in a
gap of the essential spectrum. The eigenvectors $v_\pm$ represent Jost-type
solutions:  $v_+$ represents a
solution to the periodic problem satisfying $\lim_{n \to -\infty} M^n
 v_+ = 0 $, while $v_-$ represents a solution to
the periodic problem which satisfies $\lim_{n \to +\infty} M^n v_- =0$.
In order to have an $L^2$ solution the solution which is decaying at 
$-\infty$ must connect
to the solution which is decaying at $+\infty$.
Since the defect potential is compactly supported this eigenvalue condition
reduces to the following: $N$, the transfer matrix for the defect mode,
should map $v_+$, the left Jost eigenvector for the periodic problem
to a scalar multiple of $v_-$, the right Jost eigenvector for the periodic problem.
In other words for $E \in \mathcal{G}_j \subset R_{per}$ and
$N(E)$ defined as above, the eigenvalue condition
becomes
\begin{equation}N(E)\,v_+ =\mu v_-.
\label{eig_cond}
\end{equation}
Since we are in the
 $2\times2$ case this is equivalent to the vanishing of the following inner product:
\begin{equation}\langle v_-,J\,N(E)\,v_+\rangle=f(E)=0,
\label{eig_cond_2}
\end{equation}
where $J$ is the standard skew-symmetric matrix $J=\left(\begin{array}{cc} 0 & 1 \\ -1 & 0
\end{array}\right)$ defined previously.
The function $f(E)$ is the Evans function associated to the defect potential.
Note that, by continuity $f(E)$, can be defined uniquely at the band-edge:
standard results (see Kato\cite{K}) guarantee that the eigenvectors have a
Puiseux series representation in powers of $(E-E_{j})^\frac{1}{2}$ in the neighborhood
of a band edge $E_{j}$. In a more general setting the Evans function is
defined as a determinant of exponentially decaying Jost-type solutions, but
throughout this paper we will make extensive use of the fact (obviously
unique to the second order case) that the determinant can be written as
the inner product
with the standard  skew-form $J$. This, together with the fact that the
Hamiltonian nature of the
problem respects the skew-form, will simplify many of the calculations.

The zeros of the Evans function defined in Eq. (\ref{eig_cond_2})
define the defect eigenvalues. However it is convenient
to consider the width of the defect $x$ as a parameter. In this
way we can make a homotopy argument in $x$, and count the roots of the
above equation as they are created, with $x=0$ corresponding to
a purely periodic problem and $x=1$ to the periodic problem with a
defect. If $N(E,x)$ is the fundamental solution
operator for the defect potential as defined in \eqref{eqn:fundsol} we
define a generalized Evans function to be
\begin{equation}
f(E,x) = \langle v_-\, ,JN(E,x) v_+\rangle,~~~~~~~x\in[0,1].
\label{gen_evans_fn}
\end{equation}
The goal is to apply an implicit function argument to
show that the zeros of the generalized Evans function lie on
continuous curves $E(x)$, and thus can be tracked as functions of the
homotopy parameter $x$.

Throughout this paper $\mathcal{G}_j=(E_{2j+1},E_{2j+2})$ will denote a gap in 
the essential 
spectrum, which is assumed to be non-empty. We will let $D_{\mathcal{G}_j}$ 
represent the number of defect eigenvalues in $\mathcal{G}_j$:
\[
D_{\mathcal{G}_j} = \#(E\in \mathcal{G}_j \mid f(E)=0).
\]
Of course the goal of the paper is to understand $D_{\mathcal{G}_j}$ for a general
potential. 

Finally $\mathcal{R}_{def}$ will denote the set
\[
\mathcal{R}_{def}\{ E \mid |\tr(N(1,E))|^2 \ge 4 \},
\]
the energies for which the map across the defect has real 
eigenvalues. If one considers the periodic Schr\"odinger operator where 
the potential is the defect potential $q_{def}$ which is defined on 
$[0,1]$ is extended to a $1$-periodic function 
in the obvious way, then the set $\mathcal{R}_{def}$ is the union of 
the resolvent set of this operator with the periodic and antiperiodic 
eigenvalues. Our count of the number of point eigenvalues in a 
gap can be expressed in terms of 
the number of components of the intersection of this set with the 
resolvent set of the periodic problem. Note that, despite the similarity
in notation, $\mathcal{R}_{def}$ is a closed set whereas 
$\mathcal{R}_{per}$ is an open set, for reasons that should become clear.

\subsection{Preliminary lemmas}

We start with some facts about the dependence of the eigenvalues and
eigenvectors of a $2\times2$ real symplectic matrix $M(E)$
on a real parameter $E$.
We assume that $E\in\mathcal{R}_{per}$, so that the eigenvalues
are real and distinct. The eigenvectors $v_{\pm}$, while not orthogonal,
are linearly independent and form a basis. We will normalize these vectors
so that $\Vert v_+\Vert =\Vert v_-\Vert=1$.
It is convenient to define a second set of vectors, the dual basis,
$v_+^t J^t$, $v_-^t J^t$. It is clear that these two sets are bi-orthogonal
(though typically not orthonormal).

\begin{lemma}
Suppose that  $M(E)$ is defined as in \eqref{eqn:ham1} has real distinct 
eigenvalues $|\lambda_-|<1<|\lambda_+|$ and eigenvectors $v_\pm$ ($|v_\pm|=1$), 
and define 
$\alpha_{\pm}=\langle \frac{dv_\pm}{dE},J v_\pm\rangle.$ Then we have 
\begin{equation}
\alpha_\pm = \frac{\lambda_\pm \langle v_\pm, J M_E v_\pm \rangle}{1 - \lambda_\pm^2}.  \label{eqn:alpha}
\end{equation}
Similarly if $N(E,x)$ defined in \eqref{eqn:fundsol} has real distinct 
eigenvalues 
$\tau_\pm$ with $|\tau_-|<1<|\tau_+|$ and associated eigenvectors $w_\pm$ 
($|w_\pm|=1$) and we define 
\begin{eqnarray}
&\langle \frac{dw_\pm}{dE},J w_\pm\rangle = \beta_\pm& \\
&\langle \frac{dw_\pm}{dx},J w_\pm\rangle = \delta_\pm&
\end{eqnarray}
then the functions $\beta_\pm,\delta_\pm$ satisfy
\begin{eqnarray}
&\beta_\pm = \frac{\tau_\pm \langle w_\pm, J N_E w_\pm \rangle}{1 - \tau_\pm^2}& \label{eqn:beta}\\
&\delta_\pm = \frac{\tau_\pm \langle w_\pm, J N_x w_\pm \rangle}{1 - \tau_\pm^2}\label{eqn:delta}& 
\end{eqnarray}
Note that the normalization $\Vert w_\pm\Vert=1$ implies that
\begin{align}
&\frac{d w_\pm}{dE} = \beta_\pm J w_\pm,\\
&\frac{d w_\pm}{dx} =\delta_\pm J w_\pm,
\end{align}
and similarly $v_\pm$.
\end{lemma}

\begin{proof}
The eigenvalue equation is
\begin{equation}M\,v_{+}=\lambda_{+}\,v_{+},
\label{eig:diff}
\end{equation}
Differentiation with respect to the parameter $E$ gives
\begin{equation}\frac{d M}{dE}\,v_++M\frac{d v_+}{dE}=\frac{d\lambda_+}{dE}v_{+}
+\lambda_{+}\,\frac{d v_+}{dE}.\end{equation}
Taking the second equation and multiplying by $J$ and dotting with
$v_{\pm}$, and using the fact that $M$ is symplectic, gives
\begin{eqnarray*}
\alpha_{+}(\lambda_+^{-1}-\lambda_+)&=&\langle v_{+},J\,M_E\,v_{+}\rangle
\\
\alpha_{-}(\lambda_-^{-1}-\lambda_-)&=&\langle v_{-},J\,M_E\,v_{-}\rangle
\end{eqnarray*}
\label{prop:FH}
Multiplying by $J$ and dotting with $v_\mp$ gives the equation for the
change in the eigenvalue, and similar calculations give the equations
for the other parameters. Note that all of the denominators
are non-zero away from the band edges.
\end{proof}

\begin{remark}
This is a slight modification of the Hellmann-Feynman lemma for 
self-adjoint matrices to the case of matrices in $Sp(2,{\bf R})$ with
real distinct eigenvalues.
Note that this result only uses the fact that the matrix $M$ is in
$Sp(2,{\bf R})$ with real eigenvalues, and does not use the fact
that the monodromy matrices come from a specific second order self-adjoint
eigenvalue problem.
\end{remark}

The next two results exploit the fact that the monodromy matrices are derived
from Schr\"odinger operators to show positivity/monotonicity of
various quantities defined above. The first type of result is a
Sturm oscillation type result for the energy variable $E$, and shows that
$\alpha_\pm,\delta_\pm,\beta_\pm$ are of a fixed sign. Since, by the above
calculation, these parameters represent a sense of rotation of the
eigenvectors,
this result shows that (as a function of the energy) one
eigenvector winds about the origin in a positive sense as the
energy increases, and the other winds in a negative sense.

\begin{proposition}
The monodromy matrices $M(E), N(E,x)$ satisfy the differential
equations (in $E$)
\begin{eqnarray}
M_E &=& M J \Phi(E) \\
N_E &=& N J \Theta(E,x)
\label{lem:pd}
\end{eqnarray}
where the matrix $\Phi$ is positive definite, and $\Theta(E,0)=0$ and
$\Theta$ is positive definite for $x>0$.
\end{proposition}

\begin{proof}
This calculation follows that given in  Magnus and Winkler~\cite{MW}.
It is easy to see that $N_E$ satisfies the differential equation
\begin{equation}
N_{xE} = J H N_E + J H_E N ~~~~~N_E(E,0) = 0
\end{equation}
and a solution by variation of parameters gives the formula
\begin{equation}
N_E = N\int_0^x N^{-1} J H_E N.
\end{equation}
Using the fact that $N^t J N = J$ the above becomes
\begin{equation}
N_E = N J \int_0^x N^t H_E N.
\end{equation}
It is clear that if $H_E$ is positive semi-definite then $\int N^t H_E N$
is also positive semi-definite, and is in fact positive definite for
$x>0$ unless there exists a non-zero vector $v$ independent of $y$ such
that $H_E N(E,y) v = 0$ for all $y$. For the equations considered here it is
easy to see that no such $v$ can exist, and therefore $\int_0^x N^t H_E N(E,y) dy$
is positive-definite. Similarly $M(E)$ has the representation
\begin{equation}
M_E = M J \int_0^a U^t(E,y) H_E U(E,y) dy.
\end{equation}

\end{proof}

\begin{remark}
The fact that $M_E = M J \Phi$ (and similarly $N_E$) is simply the
fact that the Lie algebra associated to the symplectic group is
matrices of the form $J H$ with $H$ symmetric. The positivity of
$\Phi$ is essentially the Sturm oscillation theorem.
\end{remark}

\begin{corollary}
Let $v_\pm, w_\pm$ be the eigenvectors in a gap for the periodic and defect
monodromy matrices respectively, and  $\alpha_\pm, \beta_\pm,\delta_\pm$ be
defined as in  Eq. (\ref{eqn:alpha},\ref{eqn:beta},\ref{eqn:delta}):
\begin{eqnarray}
\frac{\partial v_\pm}{\partial E} &=& \alpha_\pm J v_\pm \\
\frac{\partial w_\pm}{\partial E} &=& \beta_\pm J w_\pm \\
\frac{\partial w_\pm}{\partial x} &=& \delta_\pm J w_\pm
 \end{eqnarray}
then $\beta_+,\alpha_+>0$ and $\beta_-,\alpha_-<0$. Further, if the
Hamiltonian matrix in the defect region $H_{def}(E,x)$ is positive-definite,
then $\delta_+>0$ and $\delta_-<0.$
\end{corollary}

\begin{proof}
We know that the $\beta_\pm$ satisfy
\begin{eqnarray*}
\beta_{+}\,&=&\frac{\lambda_+\langle w_{+},J\,N_E\,w_{+}\rangle}{1-\lambda_+^2}
\\
\beta_{-}\,&=&\frac{\lambda_+\langle w_{-},J\,N_E\,w_{-}\rangle}{1-\lambda_-^2},
\end{eqnarray*}
Using the fact that $N_E = N J \Theta$ and $N^t J N = J$ we have that
\begin{eqnarray*}
\beta_{+}\,&=&\frac{\lambda_+\langle w_{+},J\,N \, J \Theta\,w_{+}\rangle}{1-\lambda_+^2}
\\
&=&\frac{-\lambda_+\langle w_{+},N^{-t}\Theta\,w_{+}\rangle}{1-\lambda_+^2} \\
&=&\frac{\langle w_{+},\Theta\,w_{+}\rangle}{\lambda_+^2-1} > 0
\end{eqnarray*}
and similarly $\beta_-,\alpha_+,\alpha_-$. Using the fact that $\Theta$ is positive-definite
and $|\lambda_+|^2 > 1$ we have that $\alpha_+>0,\alpha_-<0$.

The second part follows similarly: $JN_x= -H_{def} N,$ thus
\begin{equation}
\delta_\pm =\frac{\lambda_+\langle w_{+},-H_{def} N w_{+}\rangle}{1-\lambda_+^2}
\end{equation}with $H_{def}$ positive definite,
and the
above argument follows.

\end{proof}

\section{Counting the Defects}

In this section we present the main results, which allow us to estimate the
number of gap eigenvalues created by a defect potential in terms of the
winding number of the monodromy matrix at the band edge. Again the basic
idea is to make a homotopy argument in $x$ of the generalized Evans
function $f(E,x)$, and count the eigenvalues as the are created.

The first important result is the following:
\begin{lemma}
Suppose that $E \in \mathcal{R}_{def}$. The Evans function $f(E,x)$ and 
its first partial with respect to the energy $\frac{df}{dE}$ cannot
vanish simultaneously.
For energies in the classically allowed region for the defect
$\frac{\partial f}{\partial x}$ cannot vanish at a zero of the Evans
function, and has the same sign as $\frac{\partial f}{\partial E}$.
\end{lemma}

\begin{proof}
Obviously we have the following expression for $f_E:$
\begin{equation}
f_E=\langle \frac{dv_{-}}{dE},J\,N\,v_+\rangle+\langle v_{-},J\,N\,\frac{dv_+}{dE}\rangle+
\langle v_{-},J\,\frac{dN}{dE}\,v_+\rangle.
\end{equation}
Using the identities $\frac{dN}{dE}=NJ\Theta$ and $N^tJN=J$ the above expression reduces to
\begin{equation}
f_E = \alpha_- \langle v_- N v_+\rangle - \alpha_+ \langle v_- N^{-t} v_+\rangle
- \langle v_- N^{-t} \Theta v_+\rangle.
\end{equation}
We would like to compute the signs of the various terms above at a
zero of the Evans function $f$. Note that, at a zero, we have $N v_+ = \mu
v_-$, or $N^{-1} v_- = \mu^{-1} v_+.$ Using this fact we have
\begin{equation}
f_E = \mu \alpha_- - \frac{\alpha_+}{\mu} - \frac{1}{\mu} \langle v_+, \Theta v_+\rangle
\label{eqn:feder}
\end{equation}
Since $\Theta$ is positive definite, $\alpha_-<0$ and $\alpha_+>0$
we have that $f_E$ cannot vanish at a zero of the Evans function and 
has the opposite sign from $\mu$. A similar but more straightforward 
calculation gives
\begin{eqnarray}
f_x &=& \langle v_-, J N_x v_+\rangle = \langle v_-, J J H_{def} N v_+\rangle \\
&=& -\mu \langle v_-, H_{def} v_-\rangle
\end{eqnarray}
where the last step assumes that one is at a zero of $f(E,x).$
Again $H_{def}$ is positive definite in the Schr\"odinger case for energies
in the classically allowed region (for the defect) and in the other cases
without restriction.
\end{proof}

This result has a small technical drawback, in that the derivative of the
Evans function with respect to energy diverges near the band edges of the
periodic problem,
since the spectral quantities have square-root type singularities near the
band edges. These can easily be eliminated by choosing appropriate coordinates
in the gap.

\begin{lemma}\label{gaplemma}
For a given gap $\mathcal{G}_j=(E_{2j+1},E_{2j+2}) \subset \mathcal{R}_{per}$
define the new variable $\tilde E$ by the
invertible map
\[
\tilde E = \int_{E_{2j+1}}^E \frac{dE}{\sqrt{k^2(E)-4}}
\]
that maps $[E_{2j+1},E_{2j+2}]$ to $[0,\omega_j],$ with $\omega_j = \int_{E_{2j+1}}^{E_{2j+2}} \frac{dE}{\sqrt{k^2(E)-4}}$.
Then $f$ is a $C^1$ function of $\tilde E$ in $[0,\omega_j]$, and
$f$ and $f_{\tilde E}$ cannot vanish simultaneously in $[0,\omega_j]$.
\end{lemma}

\begin{proof}
This is a straightforward chain-rule argument. The eigenvalue equation
$\lambda^2 - k(E) \lambda + 1 =0$ is equivalent to 
 \[\lambda_{\pm}^2-1=\pm \lambda_{\pm} \sqrt{k^2(E)-4}.\] From this together 
with the relation
\[\frac{d \tilde E}{dE}=\frac{1}{\sqrt{k^2(E)-4}},\] 
we have that
\[ f_{\tilde E}(\tilde E)=f_E \frac{dE}{d \tilde E}=-\mu \langle M^{-1} v_-,\Phi v_-\rangle - \frac{1}{{\mu}} \langle M^{-1} v_+,\Phi v_+\rangle -
\frac{1}{\mu} \langle v_+, \Theta v_+\rangle \sqrt{k^(E)^2-4}. \]
At the band edge, it holds $k(E)^2=4$ and $M v_{\pm}= \lambda_{\pm} v_{\pm}$ with $|\lambda_{\pm}|=1$, and $v_+(0)=v_-(0)$ so it follows that
\[
\lim_{\tilde E \rightarrow 0} f_{\tilde E}(\tilde E)= -\frac{\mu^2+1}{\mu}
\langle v_+(0),\Phi(0)v_+(0)\rangle
\]
 where $v_+(0)$ is the (unique) eigenvector of $M$ at the band-edge
$\tilde E = 0$, $\mu$ is defined by $N(0,x^*) v_+(0)= \mu v_+(0)$ which holds at a zero of $f$, and
$\Phi$ is the positive-definite matrix defined in (\ref{lem:pd}). A similar
result holds for the upper band-edge $\tilde E = \omega_j$
\end{proof}

\begin{remark}
The above is the hyperelliptic integral associated to the spectral
problem\cite{MVM,BBEIM}. Note that the
integral is necessarily convergent, since the Floquet discriminant
$k(E)$ necessarily crosses $\pm 2$ transversely at the edge of a
non-empty gap.

\end{remark}

This lemma has the following obvious implication

\begin{corollary}
Suppose that the gap $\mathcal(G)=(0,\omega_j)$ is a gap 
(parameterized by $\tilde E$ as in lemma \ref{gaplemma}).
A given zero level set of the generalized Evans function
$f(\tilde E,x)$ is given by a function $\tilde E(x)$ which is either
defined for all $x\in[0,1]$, or leaves the range $(0,\omega_j)$ at some
point(s). In the classically allowed region $E>q_{def}$ this function is
strictly monotone and thus invertible, $x(\tilde E)$.
\end{corollary}

\begin{proof}
A standard implicit function argument. Local existence
is obvious. Since the Evans function and its derivative with
respect to energy cannot vanish simultaneously  we have uniform
control on the Lipschitz constant on $[0,\omega_j]\times[0, 1]$, and thus
the level set can be constructed for all $x$ as long as
 $\tilde E(x)$ remains in $[0,\omega_j].$

\end{proof}

From this it should be clear that in the classically allowed
region for the defect the number of roots of the
Evans function $f(\tilde E,1)$ can be expressed in terms of the number
of roots of the Evans function at the band edges $0$ and $1$.
This is the content of the next proposition.

\begin{proposition}
Suppose that the gap $(0,\omega_j)$ is in the classically
allowed region for the defect potential.
Let $N$ be the number of roots of $f(\tilde E, 1)$ in the gap
$\tilde E \in (0,1)$
$n_1$ be the number of roots of $f(0, x)$ for $x \in (0,1]$
and $n_2$
be the number of roots of  $f(\omega_j, x)$ for $x\in (0,1).$ Then $N=n_2-n_1+1.$
\end{proposition}

\begin{proof}
This is a fairly standard lemma - that simple roots persist under a homotopy
that fixes the endpoints.

Consider the region $(\tilde E,x) \in [0,\omega_j] \times [0,1]$. It is easy to see that
for $x=0$ the Evans function $f(\tilde E, 0)$ is nonzero for $\tilde E\in
(0,\omega_j)$ and vanishes at the boundaries $\tilde E = \{0,\omega_j\}$, since
the monodromy map $N(E,x)$ is the identity for $x=0$ and 
$v_\pm$ are non-degenerate on the interior of the band and degenerate
at the band edges. Since $x(E)$ is decreasing the zero of $f(\tilde E,x)$
at $(\tilde E=0,x=0)$ does not extend into the interior. If $f(\tilde E,x)$
vanishes at the band edge $(\omega_j,1)$ this zero also does not extend
into the interior.
By the above lemmas all other zeroes of the generalized Evans function can be
continued to
a zero level-set $x(\tilde E)$ that is a monotone decreasing
function.  Thus, each zero of the generalized Evans function
on the right-hand boundary $\tilde E = \omega_j$ extends to a
unique curve that must intersect the
left-hand boundary $\tilde E = 0$ or the top boundary $x=1$.
Thus we have $n_2 + 1 = N + n_1.$

\end{proof}

The first homotopy argument shows that the number
of roots of the Evans function (in $\tilde E$) is equal to the
difference in the number of roots (in $x$) on the band edges of the
generalized Evans
function. We would next like to
be able to estimate the number of roots of the generalized Evans
function in terms of the band-gap structure of the defect fundamental solution matrix.
The next lemma shows that zeros of the Evans function on the
boundary can only occur in gaps or band-edges for the
fundamental matrix $N(\tilde E,x)$ for the defect mode, but never in
bands.

\begin{lemma}
\label{lem:xgap}
Suppose that the generalized Evans function $f(\tilde E,x)$ vanishes
for an energy at a band-edge of the essential spectrum, $(\tilde E,x)=(0, x^*)$ or
$(\tilde E,x)=(\omega_j,x^*)$. 
Then $k^2_{def}(\tilde E,x^*)\ge 4.$ 
\end{lemma}

\begin{proof}
If $\tilde E=0$ or $\tilde E = \omega_j$ the eigenvectors
$v_+$ and $v_-$ are linearly dependent, and thus the condition that the
Evans function implies that
\[
f(\tilde E, x)=0 \Leftrightarrow \langle v_+,J N v_+\rangle = 0  \Leftrightarrow
N v_+ = \mu v_+.
\] Thus $v_+$ is an eigenvector of
$N(\tilde E, x^*).$ Since $v_+$ is a real vector it can be an eigenvector
of $N$ only if $N$ has real eigenvalues, which is equivalent to 
$k^2_{def}(\tilde E,x^*)\ge 4$.
\end{proof}

The above shows that zeros of the Evans function on the boundary
can only occur in gaps of $N$. We would like to show that exactly
one zero of the Evans function emerges from each gap.
The next lemma, which will be needed
to show this, says that gaps in $x$ behave qualitatively like gaps
in $\tilde E$ for energies in the classically allowed region, in the sense that
in an open gap the Floquet discriminant
necessarily crosses $\pm 2$ transversely.
The analogous
statement for gaps in $E$ is standard (see Magnus and Winkler)
and will not be proven here.

\begin{lemma}
Suppose that $x_0$ is a periodic or anti-periodic point for the defect,
$k_{def}(E,x_0)=\pm 2$
and that the energy $E$ is in the classically allowed region for the defect. 
Then either
the Floquet discriminant
crosses transversely, $\frac{dk_{def}}{dx}(E,x_0)\neq 0$, or 
$x_0$ is a double point
$N(E,x_0)=\pm I$ and $k_{def}=\pm (2- c(x-x_0))^2 + o((x-x_0)^2))$ with $c>0.$
In particular for an open gap the eigenvectors are degenerate only at the
band edges.
\end{lemma}

\begin{proof}
We prove for the case $k_{def}(E,x_0)=2$. The case  $k_{def}(E,x_0)=-2$ follows 
similarly.
If  $k_{def}(E,x_0)=2$ then $N$ takes the Jordan normal form
$N = I + \kappa~ v \otimes J v$ where $v$ is the (non-zero) eigenvector of $N$.
Note that $\kappa$ can vanish, in which case $N$ is the identity matrix.
Taking the identity $\tr(N^2)-(\tr(N))^2 = -2\det(N)=-2$, differentiating and
substituting into the Jordan normal form
gives $\frac{dk_{def}}{dx} = \kappa <v J N_x v>=-\kappa <v H_{def} v>.$ Since 
the energy is in the classically allowed region for the defect the  Hamiltonian
is positive definite and the vanishing of $\tr(N_x)$ implies that $\kappa=0$
and thus that the Jordan normal form is the identity. In this case we find
that $\frac{d^2k_{def}}{dx^2} = \tr(JH(E, x_0)JH(E, x_0)) = -2 \det(H(E,x_0))<0.$
\end{proof}

The next lemma shows that the Evans function has exactly one
zero in a gap (in $x$) of $N(\tilde E, x)$ if the
Hamiltonian $H_{def}$ is positive-definite.

\begin{lemma}
Suppose $\tilde E$ is a fixed energy such that
$H_{def}(E, x)$ is positive-definite, $v_+$ a fixed real vector,
and $(x_{low},x_{high})$ is an interval such that
\begin{itemize}
\item $k^2_{def}(\tilde E,x_{low}) = 4 = k^2_{def}(\tilde E,x_{high})$
\item $k^2_{def}(\tilde E,x) > 4$ for $x \in (x_{low},x_{high}).$
\end{itemize}
In other words $(x_{low},x_{high})$ is a gap in x of 
$N(\tilde E, x)$.
Then the condition
\[
\langle v_+ J N(\tilde E, x) v_+\rangle = 0
\]
has exactly one root in $x \in [x_{low},x_{high}].$
\end{lemma}

\begin{proof} This follows from a simple monotonicity argument. We assume
that the gap is not a double point, in which case the lemma is
trivially true. The eigenvectors of $N(\tilde E, x)$ are degenerate at the
band edges and (by the previous lemma) non-degenerate
in the band interior. We can
choose the eigenvectors $w_\pm$ in such a way that $w_+=w_-$ at the lower
band edge $x_{low}$. From Corollary 1 it follows that $w_+$ rotates
clockwise and
$w_-$ rotates counterclockwise until $w_+=-w_-$, which (by the previous
lemma) occurs at the upper band edge.
By continuity it follows that the angle between $w_+$ and $w_-$
goes through $\pi$ radians.
The eigenvalue condition is equivalent to either $w_+=v_+$ or
$w_-=v_+.$ It follows that {\em either} $v_+$ is equal to a band-edge
eigenvector of $N(\tilde E, x)$, in which case the above eigenvalue condition
has a simple root at the appropriate band edge, {\em or} $v_+$ is not a band edge
eigenvector, in which case there is exactly one root of either
$w_+=v_+$ or $w_- = v_+$ (but not both). This is illustrated in Fig. (1).
\end{proof}

Lemma \ref{lem:xgap} showed that the eigenvalue condition on the 
boundary of a gap in the essential spectrum could only have a 
roots in the set $k^2_{def}(E,x)\ge 4.$ The above lemma shows that the 
eigenvalue equation has exactly one root in each interval of this form.

\begin{figure}
\begin{center}
\begin{tabular}{c}
\psfig{file=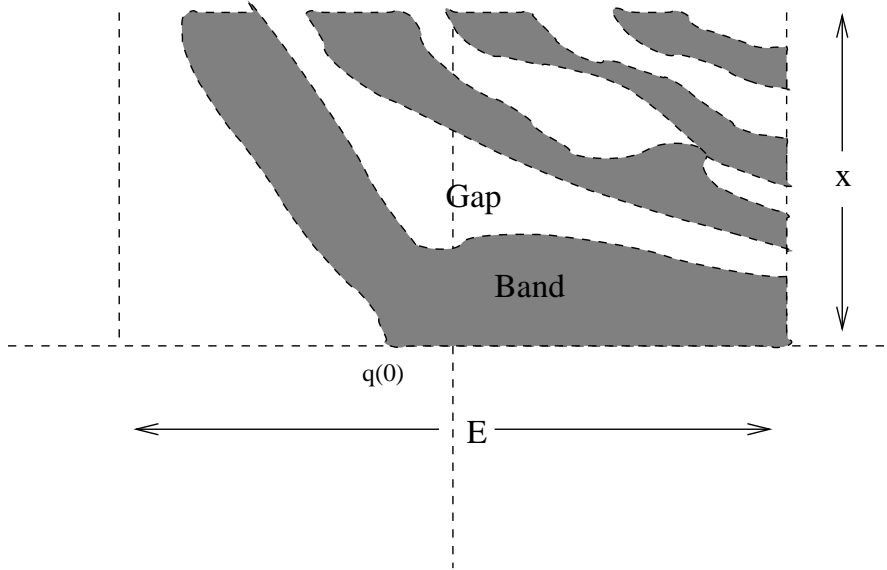,height=3.0in}
\end{tabular}
\end{center}
\caption{The picture of the band-gaps in the $(E,x)$ plane. The bands can
touch at a double point, but cannot cross. Gaps, however, can disappear and
reappear. }
\label{fig:bandgap}
\end{figure}

%\begin{figure}
%\begin{center}
%\begin{tabular}{c}
%\psfig{file=bandtest.eps,height=3.0in}
%\end{tabular}
%\end{center}
%\caption{ The behavior of the defect modes in the $(E,x)$ plane. The red lines
%represent the defect eigenvalues. Since the defect potential is constant the
%gaps in the periodic defect spectrum are closed to double points. The bands
%of the periodic defect spectrum are colored according to Krein sign, with
%green
%representing positive Krein sign and Blue representing negative. The
%transition
%between the two marks the double point. }
%\label{fig:defmode}
%\end{figure}

At this point we have shown that, for energies in the classically 
allowed region for the defect, eigenvalues persist under 
homotopy in $x$. We have also shown that there are certain distinguished 
intervals in $x$, $[x_{low},x_{high}]$, where eigenvalues can enter the gap 
through the band edge $E=E_{2j+2}$ or leave the gap through the band edge 
through the band edge $E=E_{2j+1}.$ Further we have exactly one eigenvalue 
entering or leaving the gap in each such interval. Thus the 
number of eigenvalues in the gap is equal to number which enter through
$E=E_{2j+2}$ minus the number which leave through $E_{2j+1}$. 
This is not a particularly convenient charactertization, since it 
requires us to count intervals in the homotopy parameter $x$. 
Thus we give a second homotopy argument to count the
number of gaps (in $x$) of $N(E,x)$ in terms of the number of
gaps (in $E$) for $N(E, 1)$.

\begin{lemma}
For $\delta$ sufficiently small the band gap structure of
$N(E, \delta)$ in $E$ for $E$ in a fixed interval about $q(0)$
consists of
\begin{itemize}
\item Gap for $E<q(0) - o(1)$
\item Band for $E>q(0) + o(1)$
\end{itemize}
Further there is a unique $E$ in $(q(0)-o(1),q(0)+o(1))$ where $\tr(N(E, \delta)=2$, and
the crossing is transverse.

\end{lemma}

\begin{proof}
This result follows from Taylor expansion, self adjointness, and
some elementary analytical considerations.
First note that $N(E, 0)=I$. Differentiating gives $\tr N_{xx}(E, 0) = -2 (E-q(0))$,
$\tr N_{xE}(E, 0)=0$ and $\tr N_{EE}(E, 0)=0.$ So for small $x$
$k(E,x)=\tr N(E,x)\approx 2-(E-q(0))x^2 + O(x^3).$ Therefore, it is
clear that for $x$ sufficiently small, $k(E,x)> 2$ for $E<q(0)-O(\delta)$ and $k(E,x)<2$ for
$E>q(0)+O(\delta).$ Next we need to show that there exists a
single transverse crossing in $(q(0)-O(\delta),q(0)+O(\delta)).$  To do this
note that $N(E,x)$ is the monodromy map for a self-adjoint operator, and
standard results (see Magnus and Winkler) show that the only possible band
edges are a proper band edge, which crosses transversely (in this case
$k(E,x)=2, k_E(E,x)<0$) or a double point (in this case
$k(E,x)=2, k_E(E,x)=0, k_{EE}(E,x) < 0$). Thus the first crossing must be
transverse. To show that this crossing is unique for
sufficiently small $\delta$ we note that self-adjointness implies that the
Floquet discriminant cannot have a critical point inside of a band. Suppose
that there were two crossings in $(q(0)-O(\delta),q(0)+O(\delta))$. Then,
since critical points inside of a band are disallowed, there must be a gap
in $(q(0)-O(\delta),q(0)+O(\delta))$,
and thus two antiperiodic points. Thus the Floquet discriminant must go
from $+2$ to $-2$ in a distance of $O(\delta)$, and we have an estimate of
the form $|k(E_1, \delta) - k(E_1+\delta, \delta)| \ge O(\delta^{-1}).$ Since
the Floquet discriminant is continuous this cannot hold for $\delta$
arbitrarily small, and thus the crossing is unique for small $\delta$.
\end{proof}

Now we are in a position to prove the next result:

\begin{proposition}
Suppose that $E$ is in the classically
allowed region for the defect mode.
The number of bands of $N(E, x)$ in $x\in[0,1]$ is the equal to the number of
bands of $N(E, 1)$ for $E \in (-\infty, E]$. The number of gaps of  $N(E, x)$ in $x\in[0,1]$ is one less than the number of gaps of $N(E, 1)$ for $E \in (-\infty, E]$.
\end{proposition}

\begin{proof}
This again makes use of a homotopy argument. The basic observation is that, 
while the gaps can collapse to a point and thus cannot be continued by an 
implicit function argument the bands can be so continued. 

It is somewhat more convenient to
work on the the interval $[\delta,1]\times[e,E]$ to avoid trouble with
noncompactness in $\tilde E$ and the degeneracy at $x=0$.
Since $q_{def}$ is assumed to be piecewise $C^1$ we choose
$e < inf_{x\in[0,1]} q(x).$
In this case it is clear from the obvious lower bounds that there are no bands
for energies less than or equal to $e$. From the above lemma we can
choose $\delta$
sufficiently small such that $E<q(0) - o(1)$ is a gap, $E>q(0)+o(1)$ is a band,
and there is a single periodic point at $x=\delta, E=q(0)+o(1)$. Since the bands are
by definition open we can chose $\delta$ such that the number of bands for $x \in [0,1]$ is
the same as the number of bands for $x \in [\delta,1].$

We define the bottom boundary  $B$ to be $[e, \bar E]\times \delta$, the top
boundary $T$ to be  $[e,E) \times 1$, the right boundary $R$ to be
$E \times [\delta,1]$,
and the left boundary $B$ to be $e \times [\delta,1] $.  Since $\frac{dk}{dE}$
 cannot vanish on the interior of a band it follows from the implicit function
theorem that a point in a band defines a local level set of the
the Floquet discriminant $E(x)$. Since we have global control of the
Lipschitz constant this level set is defined for all $x$.

By construction the left boundary $L$ is chosen to be below the 
essential spectrum and has no bands. The bottom boundary
$B$ has a single band $(q(0)+O(\delta),E)$, and by construction
$d k/dx<0$ on this band. This band continues onto the right boundary $R$. 
Since $E$ is in
the classically allowed region
$d k/dx<0$ for all bands on top boundary $R$. Thus all bands of the
left and top boundaries move into the interior transversely as $x$
increases.  Since a level set of $k(E,x)$ on the interior of a band
is defined globally as a function of $x$ (again denoted $E(x)$) each point on
a band interior on
the top boundary moves into the interior as $x$ is increased, and
can be followed until it intersects the right boundary. This defines a map of
bands on the right boundary to bands on the top boundary.
The converse
argument defines the inverse map of bands on the top boundary to bands on the 
right boundary.
Since we have an invertible map of band on the top boundary to bands on the
right boundary the number of bands on the top boundary and
on the right boundary must clearly be the same. Since the bands are open 
and disjoint the complements must also have the same number of components.
The bottom boundary contains one gap, the semi-infinite one, and the 
remainder of the gaps occur on the right boundary, whence the gap count.

\end{proof}

\begin{remark}
Note that the fact that the derivative $\partial k/ \partial x$ is of the same
sign as
$\partial k/\partial E$ is critical to this argument, and depends crucially on
being in the classically allowed region for the defect. Without control
on the sign of  $\partial k/\partial x$ (or equivalently $dx/dE$) there
is no guarantee that a band could not enter and leave the region repeatedly.
Physically this seems to be a non-resonance condition.
\end{remark}

From this it follows that we have the following bounds on
 the number of defect eigenvalues.

\begin{theorem}
\label{thm:eigcount}
Suppose that a gap $\mathcal{G}$ is in the classically
allowed region for the defect mode. Define the set
\[
\mathcal{R}_{def} = \{ E \vert (\tr(M(E)))^2 \ge 4 \}
\]
and the integers $n_{\mathcal{G}}$ and $n_{\partial \mathcal{G}}$ to be the number 
of connected components of the sets $\mathcal{R}_{def}\cap \mathcal{G}$ and  
$\mathcal{R}_{def}\cap \partial\mathcal{G}$ respectively.
Then, the number of defect modes $D_{\mathcal{G}}$, satisfies
the inequality 
\begin{equation}
n_{\mathcal{G}}+1-n_{\partial G} \le D_{\mathcal{G}} \le n_{\mathcal{G}}+1.\label{Top_Bounds}
\end{equation}

\end{theorem}

\begin{proof}
We'll prove the case $n_{\partial \mathcal{G}}=0$ first, in which case the count is 
exact.
If $n_{\partial G}=0$, i.e., neither edge lies in a gap for the defect monodromy
matrix, then from the previous proposition and lemma 8 it follows that
the  number of gaps of $N(\tilde E, x)$ in $x$ is the equal to the number of
gaps of $N(0, x)$ in $x$ for $x \in [0,1]$ plus $n_{G}$.
But, since in each gap (in $x$) there is a unique eigenvalue, we obtain that
(in the terminology of Proposition 1) $n_2-n_1=n_{G}$, and therefore 
$D_{G}=n_{G}+1$.

On the other hand, for each of the edges $0, 1$ that lie in a gap for the defect
monodromy, the defect modes can be reduced by $1$. This is true, since the zero level
curves for the Evans function might not exit the $x-$ interval $[0,1)$ in the case where
$E_{low}$ lies in a gap, or the defect mode might be generated
outside of $[0,1)$ in the  case where $E_{high}$ lies in a gap.
\end{proof}

\begin{remark}
It is worth making a number of remarks about this result. First we note that 
one could easily find upper and lower bounds on the number of defect eigenvalues
generated in a gap by counting Dirichlet eigenvalues of the defect. However 
using this method there seems to be no clear criterion for guaranteeing that 
the count is exact, which will be necessary for the main result of this paper,
the large gap number estimate. 

As noted earlier the set $R_{def}$ is the union of the resolvent set and the 
periodic and anti-periodic eigenvalues of the periodic Schr\"dinger operator 
with potential given by the periodization of the defect potential. Thus 
the above result, roughly speaking, gives an estimate of the number of defect 
eigenvalues in terms of the number of connected components of the 
intersection of the resolvent sets. 

The above calculation can also be interpreted as a Maslov index calculation. 
Given a curve $N(E)$ in the symplectic group and a Lagrangian subspace 
$\mathcal{L}$ (in this case any one dimensional subspace) 
the Maslov index is a signed count of the number of 
intersections of $N(E) \mathcal{L}$ with $\mathcal{L}.$ In our case 
lemma \ref{} guarantees that the index is always positive, so the
Maslov index actually counts intersections. This count 
depends, of course, on the Lagrangian subspace chosen but in the special
case $n_{\partial \mathcal{G}}=0$ the Maslov index $\mu_{\mathcal{L}}$is the same for all 
Lagrangian subspaces and the number of defect eigenvalues is one plus the 
Maslov index of $N(E)$ for $E \in (E_{2j+1},E_{2j+2})$.   

Finally we remark that, in the case of defect potentials which are constant 
the set $\mathcal{R}_{def}$ consists of single points. In this case it 
is easy to see that the count is exactly $n_{\mathcal{G}}+1$ regardless of 
whether or not $n_{\partial\mathcal{G}}=0.$

\end{remark}

\subsection{Large gap number asymptotics}

In this section we present results for the high gap-number behavior of the
number of defect eigenvalues. It is known \cite{east} that the $n$-th Dirichlet
eigenvalue for the problem
\[
-u_{xx}+V(x) u=\lambda_{n} u,~~~~u(0)=u(L)=0
\]
behaves asymptotically as
\[ 
\lambda_{n}\sim \frac{n^2 \pi^2}{L^2}+\frac{1}{L}\int_0^L V(x) dx + o(1).
\]
We have established previously bounds for the number of eigenvalues in each
gap. In particular we have that if the gaps of the (periodically extended)
defect problem do not intersect a gap of the periodic problem, then
that gap has exactly one defect eigenvalue. Further we know the following standard facts:
\begin{itemize}
\item The $n^{th}$ Dirichlet eigenvalue is contained in the $n^{th}$ gap.
\item The width of the $n^{th}$ gap goes to zero for large $n$. In particular
if $V\in L_2$ then the sequence of widths is in $l_2$, with stronger
decay estimates if $V$ has additional smoothness properties.
\end{itemize}
Thus, if we denote the $n$-th Dirichlet eigenvalue of the periodic
problem by $\mu_n$ and the $m$-th Dirichlet eigenvalue for the defect problem
by $\tilde \mu_m$ then we
can expect to have one eigenvalue in each gap, unless $\mu_n \approx \tilde{\mu}_m$,
for some integers $m, n$. To quantify this we first define the following
set of exceptional energies, which detects possible overlaps between the
Dirichlet spectra of the periodic potential and the defect potential:

\begin{definition}
For a real number $a$ we define the set $\mathcal{F}_a$
as follows: a number $y$ belongs to $\mathcal{F}_a$ if for every $\delta >0$ there
exists an infinite, strictly increasing sequence of pairs of integers
$\{N_k,M_k\}_{k=1}^\infty$
such that
\[
|y-M_k(N_k - M_k a)|<\delta
\]
\end{definition}
In other words, the set $\mathcal{F}_a$ represents the (appropriately scaled) 
asymptotic distribution of errors in rational approximations of $a$. 
In the next lemma we note some simple
properties of the set $\mathcal{F}_a$.
\begin{lemma}

The set $\mathcal{F}_a$ has the following properties:
\begin{itemize}
\item The set $\mathcal{F}_a$ is never empty. If $a$ is rational
then $\mathcal{F}_a$ contains only the point $y=0$.
\item If $a$ is irrational then there are
at least countably many points in $\mathcal{F}_a$.
\item The set $\mathcal{F}_a$ is invariant
under the modular group 
$\PSL(2,{\bf Z})$: if $y\in \mathcal{F}_a$ with $M_k(N_k - a M_k)\rightarrow y$
and $a=\frac{n_1 b+n_2}{n_3 b + n_4}$ with $n_1 n_4 - n_2 n_3 =1$ then
$y \in F_b.$
\end{itemize}
\end{lemma}
\begin{proof}
The first property is obvious, since it is easy to see that a given
non-zero integer can be written as a difference of squares in at
most a finite number of ways.  For the irrational case note that
from elementary number theory if a is irrational there exists an
infinite sequence of pairs $(N,M)$ such that $|\frac{N}{M}-a| < \frac{1}{M^2}$,thus an
infinite sequence of pairs $(N,M)$ such that $M(N - M a)\in(-1,1)$, and
by compactness a limit point. Note that if
$y\in \mathcal{F}_a$ then $j^2 y \in \mathcal{F}_a$ for all integers $j$, so the number of
points in $\mathcal{F}_a$ is at least countably infinite.

Finally to see the invariance of $\mathcal{F}_a$ under the modular group note that
a straightforward calculation shows that
if the sequence $(N_k,M_k)$ satisfies $ M_k(N_k - a M_k)\rightarrow y$
then the sequence $N^\prime_k = n_4 N_k - n_2 M_k, M_k^\prime = n_1 M_k - n_3 N_k$
satisfies $M_k^\prime (N_k^\prime - b M_k^\prime)\rightarrow y$
\end{proof}

Finally, we characterize the set $\mathcal{F}_a$ for some classes of real numbers $a.$
\begin{proposition}
Suppose $a$ has the continued fraction expansion $a=[a_0,a_1,a_2\ldots].$
\begin{itemize}
\item If $\{a_j\}_{j=0}^\infty$ is eventually periodic (in other words if $a$ is a quadratic irrational) then $\mathcal{F}_a$ is a discrete set of point.
\item If $\{a_{2j}\}^\infty_{j=0}$ is unbounded then $\mathcal{F}_a$ contains the
negative half-line $(-\infty,0]$.
\item If $\{a_{2j+1}\}_{j=0}^\infty$ is unbounded
then $\mathcal{F}_a$ contains the positive half-line $[0,\infty)$.
\item If $\{a_j\}_{j=1}^\infty$ is
bounded then $\mathcal{F}_a$ does {\bf not} contain some interval about the origin.
\end{itemize}
In particular for Lesbesgue almost every $a$ both the even and the odd
terms in the continued fraction expansion are unbounded and thus we have
$\mathcal{F}_a =R$, the whole real line.
\end{proposition}

\begin{proof}
In the case of quadratic irrationals it is relatively easy to compute
explicitly what $\mathcal{F}_a$ is. Let $f(x)=n_1x^2+n_2x+n_3$ with
$n_{1,2,3} \in {\mathbb Z}$ relatively prime be the quadratic polynomial
with root $a$. Then $\mathcal{F}_a$ consists of all numbers of the form
$\frac{j}{ f'(a)}=\frac{j}{\sqrt{n_2^2-4n_1n_3}}$, where $j$ is any integer
which can be represented in the form $ n_1 N^2 + n_2 N M+ n_3 M^2=j.$

To see this note that standard results in the theory of quadratic Diophantine
equations show that the existence of one solution guarantees the existence of
a family of solutions.
Let $N_k, M_k$ be increasing sequences of integers such that
$N_k / M_k \rightarrow a.$ Obviously $n_1 N_k^2 + n_2 N_k M_k + n_3 M_k^2 = j$
for some integer $j.$
Since $f(a)=0$, it follows by the mean value theorem that
\[
f(a)-f(N_k/M_k)=f'(x_0) (a-N_k/M_k),
\]
for some $x_0$ between $a$ and $N_k/M_k$, and so we obtain $f'(x_0) (a-N_k/M_k)=-j/M_k^2.$
Therefore
\[M_k(N_k-a M_k)= \frac{j_k}{f'(x_0)}
\]
for $x_0\in(a, N_k/M_k)$. Thus all elements of $\mathcal{F}_a$ are of the stated form.
To see that all such numbers are actually arise note that the existence
of one solution to $ n_1 N^2 + n_2 N M+ n_3 M^2=j$ implies the existence of
a family of such solutions, which by the above must satisfy
\[
M_k(N_k-a M_k)= \frac{j}{f'(x_0)}
\]

To see the second and third claims we consider the following doubly indexed
sequence: $N_{j,k} = j P_k, M_{j,k} = j Q_k$, where $\frac{P_k}{Q_k}$ are the
continued fraction approximants to $a$. Standard results show that
$|P_k^2 - a^2 Q_k^2| \leq \frac{C}{\alpha_k}$, so if $\alpha_k$ is unbounded
there exists a subsequence such that $P_k^2-Q_k^2\alpha^2= \epsilon_k \to 0.$
Further $\epsilon_k$ is negative (resp. positive) if $k$ is even (resp. odd).
Note that for a fixed integer $j$ we have that $j^2P_k^2 - \alpha^2 j^2Q_k^2 = j^2\epsilon_k.$
It is clear that for any $y$ having the same sign as $\epsilon_k$
if we take $j_k = \lfloor\frac{y}{\epsilon_k}\rfloor$ we have
$|j_k^2 \epsilon_k-y|\leq O(\epsilon_k^\frac{1}{2}).$

The fourth observation follows from the well-known fact that
a number with a bounded continued fraction cannot be approximated by
rationals to better than quadratic order:
if the continued fraction coefficients satisfy $a_i<m$ then one
has a lower bound of the form
\[
|N/M - a | \geq f(m)/M^2
\]
with $f(m)>0,$ from which it follows that the interval $|y|<f(m)$ 
is not in the set $\mathcal{F}_a.$

The last assertion follows from a trivial modification of the proof in
Hardy and Wright that the set of numbers with bounded continued fraction
coefficients has measure zero.
\end{proof}

\begin{remark}
The case which is still incompletely understood is that for
which the continued fraction
expansion has bounded coefficients but is not periodic. It would be
interesting, although likely very difficult, to classify $\mathcal{F}_a$ based on the
distribution of the continued fraction coefficients.
\end{remark}

Now we are in a position to state the main result, which relates the asymptotic
number of eigenvalues in a gap to the properties of the set $\mathcal{F}_a.$

\begin{theorem}
Assume that the width of the defect is normalized to $1$, and as before denote
the period of the periodic potential by $a$.
Define the energy difference $\Delta q$ to be the mean of periodic potential minus the mean
of the defect potential:
\[
\Delta q = \int_0^1 q_{def}(y) dy - \frac{1}{a} \int_0^a q_{per}(y) dy.
\]
Then:
\begin{itemize}
\item If the energy difference does not belong to the set of exceptional energies, 
$\frac{a \Delta q}{2\pi^2}\notin \mathcal{F}_a,$ then
every sufficiently large numbered gap contains {\em exactly} one defect
eigenvalue.
\item If $\frac{a \Delta q}{2\pi^2}\in \mathcal{F}_a$ then there exists a sequence of exceptional gaps.
Every sufficiently large numbered gap which is not in the sequence of
exceptional gaps contains exactly one eigenvalue. Every sufficiently large
numbered gap which is in the sequence of exceptional gaps contains
$0,1,$ or $2$ defect eigenvalues.
\end{itemize}
\end{theorem}
\begin{proof}
Denote the $n$-th Dirichlet eigenvalue of the periodic
problem by $\mu_n$ and the $m$-th Dirichlet eigenvalue for the defect problem
by $\tilde \mu_m$. Then,
\begin{align}
\mu_{n}-\tilde \mu_{m} &\sim \frac{n^2 \pi^2}{a^2}-m^2 \pi^2-\Delta q+ o(1) \\
& \sim \frac{1}{a^2\pi^2} (n+am)(n-am) - \Delta q + o(1).
\end{align}
For $n,m$ large this is clearly bounded away from zero unless $n=am+o(1),$
in which case this becomes
\begin{align}
\mu_{n}-\tilde \mu_{m} &\sim \frac{2 \pi^2}{a} m(n-am) -\Delta q + o(1).
\end{align}
If $\Delta q$ does not belong to  the set of exceptional energies $\mathcal{F}_a$ then
this quantity is eventually uniformly bounded away from zero. Thus all 
sufficiently
large Dirichlet eigenvalues of the periodic problem and the defect
problem are bounded away from each other. Since the Dirichlet eigenvalues are
contained in the gaps, and the width of the gaps is approaching zero, it
follows that the gaps are eventually non-intersecting. From the count in 
Theorem \ref{thm:eigcount} it follows that 
each gap contains exactly one eigenvalue.

If $\Delta E\in \mathcal{F}_a$ then it is clear that at most one of the gaps of the
defect problem can intersect a gap of the periodic problem. Hence
the gap of contains either $0, 1$ or $2$ defect eigenvalues
according to Theorem \ref{thm:eigcount}.

\end{proof}

To close we construct an example to show that the the exceptions considered in
the previous theorem are real. In particular we construct a 
combination of periodic and defect potentials such that
a particular (infinite) subsequence of gaps contains two defect eigenvalues.

\begin{example}
In order to construct this example we would like to choose the periodic and 
defect potentials in such a way that the gaps in the essential spectrum of the 
periodic problem are comparatively wide.
For this reason we choose a piecewise constant (Kronig-Penney) 
potential. We choose the period to be a quadratic irrational since this 
guarantees that we have a clean description of the set $\mathcal{F}_a$. For reasons of tradition we take $a=\phi$ the golden mean, although any quadratic 
irrational would serve. We take the defect potential to be constant, 
$q_{def}(x)=q_{def}$, so the 
set $\mathcal{R}_{def}$ consists of a union of points, and the eigenvalue 
count is guaranteed to be exact. In particular for the periodic 
potential we take the Kronig-Penney potential
\begin{eqnarray*}
q_{per}(x)=\left\{ \begin{array}{ll}
-A& \mbox{for $0 \le x < \phi/2$}\\
A& \mbox{for $\phi/2 \le x < \phi$}
\end{array}\right.
\end{eqnarray*}
where $A$ is a constant to be determined later. The gaps grow in size with 
$A$, and our strategy is to choose $A$ sufficiently large so that appropriate 
points of  $\mathcal{R}_{def}$ are eventually contained in the appropriate gaps, 
as this will guarantee two eigenvalues.

The transfer matrix for $-u_{xx}+q_{per}(x) u= E u$ can be easily calculated 
to be 
\begin{eqnarray*}
M_{KP}=\left( \begin{array}{cc}
\cos(\frac{\phi \sqrt{E-A}}{2})& \frac{\sin(\frac{\phi \sqrt{E-A}}{2})}{\sqrt{E-A}}\\
-\frac{\sin(\frac{\phi \sqrt{E-A}}{2})}{\sqrt{E-A}}& \cos(\frac{\phi \sqrt{E-A}}{2})
\end{array}\right) \left( \begin{array}{cc}
\cos(\frac{\phi \sqrt{E+A}}{2})& \frac{\sin(\frac{\phi \sqrt{E+A}}{2})}{\sqrt{E+A}}\\
-\frac{\sin(\frac{\phi \sqrt{E+A}}{2})}{\sqrt{E+A}}& \cos(\frac{\phi \sqrt{E+A}}{2})
\end{array}\right) .
\end{eqnarray*}
From this, the location of the band-edges is given by roots of the equation
\begin{eqnarray*}
& \tr(M_{KP})=
2 \cos(\frac{\phi \sqrt{E-A}}{2}) \cos(\frac{\phi \sqrt{E+A}}{2})&\\
&-\left(\frac{\sqrt{E-A}}{\sqrt{E+A}}+\frac{\sqrt{E+A}}{\sqrt{E-A}}\right)
\sin(\frac{\phi \sqrt{E+A}}{2}) \times \sin(\frac{\phi \sqrt{E-A}}{2})=\pm 2 & .
\end{eqnarray*}

From some straightforward asymptotic analysis it can be seen that the
location of the $j^{th}$ gap is approximately
\begin{eqnarray}
\mathcal{G}_j=\left\{ \begin{array}{ll}
\left( \frac{j^2 \pi^2}{\phi^2}- \frac{\phi A}{2\pi j} + O(j^{-2}),
\frac{j^2 \pi^2}{\phi^2} + \frac{\phi A}{2\pi j} + O(j^{-2})\right) & \mbox{for $j\gg 1$ odd} \\
\left(\frac{j^2 \pi^2}{\phi^2} - \frac{\phi^2 A^2 }{4 \pi^2 j^2} + O(j^{-3}),
\frac{j^2 \pi^2}{\phi^2}+\frac{3\phi^2 A^2 }{4 \pi^2 j^2} + O(j^{-3})\right) & \mbox{for $j\gg 1$ even} \label{evengap}
\end{array}\right.
\end{eqnarray}
Since the  defect potential is constant the periodization is constant.
Thus the gaps close to double points and the set $\mathcal{R}_{def}$
is given by 
\[
\mathcal{R}_{def} = \cup_{j=0}^\infty \mu_j ~~~~~~~~~\mu_j = \pi^2 j^2 + q_{def}.
\]

The set $\mathcal{F}_\phi$ consists of real numbers of the form
\begin{equation}
\mathcal{F}_\phi  = \left\{ \frac{j}{\sqrt{5}} \right\}~~~~~~~
j  \in \pm \left\{1,4,5,9,11,16,19,20,25,29,31,36...\right\}
\end{equation}
where the sequence is all positive integers $j$ which are representable in the
form $j=n^2 - n m - m^2$ (sequence  A031363 in Sloane's encyclopedia\cite{Sloane}). The mean of the periodic 
potential is zero,
so if the mean of the defect potential is chosen to be in the set
$\frac{2 \pi^2}{\phi} \mathcal{F}_\phi$ then there is a infinite sequence 
of gaps which potentially have more than one
defect eigenvalue. We will, somewhat arbitrarily, focus on the 
$j=11$ term of $\mathcal{F}_\phi$. This term 
is interesting since it admits 
two distinct families of solutions to the Diophantine 
equation $n_k^2 - n_km_k - m_k^2 =11$. These two families are given by 
the following sequences:
\begin{align}
m_k &= \{1,5,14,37,97\ldots\} ~~~~~~~~~~n_k =\{4,9,23,60,\ldots\} \label{seq1}\\
m_k &= \{2,7,19,50,131\ldots\}~~~~~~~~~~~n_k =\{5,12,31,81,\ldots\}. \label{seq2}
\end{align}
Each of the sequences $n_k,m_k$ satisfies the recurrence 
$a_{k+1}=3 a_k - a_{k-1}$ and $n_k^2 - n_km_k - m_k^2 =11.$ 
It is straightforward to compute that the $n_k,m_k$ satisfy the following asymptotic relation
\[
m_k(n_k - \frac{1+\sqrt{5}}{2} m_k) = \frac{11}{\sqrt{5}} - \frac{121}{m_k^2 \sqrt{125}} + O(m_k^{-4}).
\]
(This is easiest to see if one notes that $n_k^2 - n_km_k - m_k^2 =11$ can 
be factored over $\mathbb{Q}(\sqrt{5})$ as  $(n_k-\phi m_k)(n_k + \phi^{-1} m_k) =11$ and proceed from there.)

We look at the $G_{n_k}$, the $n_k^{\rm th}$ gap in the essential spectrum with 
$n_k$ an element of one of \eqref{seq1} or \eqref{seq2}, and the 
$m_k^{\rm th}$ point in $\mathcal{R}_{def}$, which is located at 
$\pi^2 m_k^2 + q_{def}.$ The distance 
between the center-point of $G_{n_k}$ and $\pi^2 m_k^2 + q_{def}$ is given by 
\begin{eqnarray*}
\pi^2 m_k^2 + q_{def} - \frac{n_k^2 \pi^2}{\phi^2} &=& \pi^2 (m_k + \frac{n_k}{\phi})
(m_k - \frac{n_k}{\phi}) \\
&\approx& q_{def}-\frac{22 \pi^2}{\sqrt{5}\phi} + \frac{121 \pi^2}{5 \phi^2 m_k^2} + o(m_k^{-2}).\\
\end{eqnarray*}
Choosing $ q_{def}=\frac{22 \pi^2}{\sqrt{5}\phi}$ (again this is the $j=11$ 
element of $\frac{2 \pi^2}{\phi}\mathcal{F}_\phi$)causes the leading order 
terms to cancel. Given the gap asymptotics in (\ref{evengap})
it is clear that for $n_k$ odd the point $\mu_{m_k}$ is always eventually 
contained in the gap $\mathcal{G}_{n_k}$, since the width of the gaps 
$\mathcal{G}_{n_k}$ decays more slowly that the above error. For $m_k$ even 
the point $\mu_{m_k}$ will eventually be contained in the gap  
$\mathcal{G}_{n_k}$ if $A$ is chosen to be sufficiently large.
If we choose 
\[
\frac{3 \phi^2A^2}{4\pi^2n_k^2}> \frac{121 \pi^2}{5 m_k^2 \phi^2}
\]
or equivalently 
\[
A > \frac{22 \pi^2}{\sqrt{15}\phi} 
\]
then the radius of the $n_k$ gap is asymptotically larger than the distance 
between the $m_k$ point in $\mathcal{R}_{def}$ and the center of the gap, and 
thus the $m_k$ point in $\mathcal{R}_{def}$ is contained in the $n_k^{th}$ gap.
Thus $n_{\mathcal{G}}=1, n_{\partial\mathcal{G}}=0$ and the gap $\mathcal{G}_{n_k}$ contains
precisely 2 eigenvalues.

Note that this example is fairly robust.  For any positive $A$ the {\em odd}
gaps $\mathcal{G}_{n_k}$ for $n_k$ in the sequences \eqref{seq1}  or \eqref{seq2}
get exactly two eigenvalues: the lower bound on $A$ is necessary only to insure 
that the even gaps (which are asymptotically narrower) get two eigenvalues. 
Also note that the choice of $q_{def}(x)$ is not particularly important: as 
long as $q_{def}$ is smooth enough that the intervals in $\mathcal{R}_{def}$
decay at least as fast as the gaps in the essential spectrum then (for sufficiently large contrast $A$) the above example can be made to work. Similarly, as noted above, the above construction goes through for any quadratic irrational 
(in fact $\phi$ is in some sense the worst example since it is the least well 
approximated by rationals). For a general irrational $a$ and general potentials 
$q_{def}(x)$ and $q_{per}(x)$ there is a somewhat delicate interplay between the 
size of the intervals in $\mathcal{R}_{def}$ and $\mathcal{R}_{per}$ and the 
error in the rational approximations of $a$, and while it is clear that 
the sequence of extraordinary gaps can contain at most two eigenvalues it is 
difficult to determine the exact number. 

\end{example}

\section{Conclusion}

In this paper, we analyzed in detail the emergence and distribution of defect
eigenvalues in the gaps of the essential spectrum of the linear one-dimensional
Schr\"odinger equation, with a potential consisting of a periodic part plus a
compactly supported defect part. We used an Evans function technique that
reduces the problem to that of finding the zeros of an analytic
function, and by means of a homotopy argument we were able to count the
eigenvalues as they emerge from the band edges.
It is found that if a gap is in the classically allowed region for the
defect potential, then the number of defect modes in the gap can be
expressed in terms of the number of roots of the Evans function at its
band edges.
It is shown that for a gap in the classically allowed
region for the defect potential, the number of zeros of the generalized
Evans function is no larger than the number of gaps of the defect
problem in terms of the homotopy parameter $x$.
In addition, this number was found to be the same as that of the gaps
in terms of the eigenvalue parameter $E$.
As a result, bounds for the defect modes are given in terms of the gaps
of the defect monodromy matrix that intersect the gap.
Finally, we proved the following significant
generalization of Zheludev's theorem: the number of point eigenvalues in a
gap in the essential spectrum is exactly $1$ for sufficiently
large gap number unless a certain Diophantine approximation problem has solutions, in
which case there exists a subsequence of gaps containing $0,1,$ or $2$
eigenvalues. We stated some conditions under which the solvability of the Diophantine
approximation problem can be established, and we included an example where a particular
(infinite) subsequence of gaps contains two defect eigenvalues.

There are a number of interesting open questions. It would be interesting to 
understand the structure of the set $\mathcal{F}_a$ for $a$ irrational 
with bounded coefficients but not a quadratic irrational, although this 
would likely depend in a sensitive way on the distribution of the continued 
fraction coefficients. Also, since the bounds are expressed in terms of the 
intersection of resolvent sets it would be interesting to understand how 
these are effected by isospectral flows on the potential(s), which obviously 
leave these sets invariant.  
 We believe, although we have not 
yet checked this, that by flowing the periodic potential according to 
the Korteweg-DeVries hierarchy (which obviously leaves $\mathcal{R}_{per}$
invariant) one should be able to force any particular gap to 
achieve any of the possibilities allowed by \eqref{Top_Bounds}.
  
\bibliography{new_defect}

\end{document}